\documentstyle[twocolumn,pra,aps]{revtex}

\begin{document}

\bibliographystyle{prsty}

\draft

\preprint{to be submitted to Phys.~Rev.~A}

\tighten

\title{Quantum control of atomic systems by\\
       homodyne detection and feedback}

\author{Holger F. Hofmann$^{1}$, G\"unter Mahler$^{2}$, and Ortwin Hess$^{1}$}
\address{
$^{1}$ Institut f\"ur Technische Physik, DLR,
       Pfaffenwaldring 38-40, 70569 Stuttgart, Germany\\
$^{2}$ Institut f\"ur Theoretische Physik und Synergetik, 
       Pfaffenwaldring 57, 70550 Stuttgart, Germany}

\date{\today}

\maketitle

\begin{abstract}
We investigate the possibilities of preserving and manipulating the 
coherence of atomic two level systems by ideal projective homodyne detection 
and feedback. 
For this purpose, the photon emission process is described on timescales
much shorter than the lifetime of the excited state using
a model based on Wigner-Weisskopf theory. The back-action of this 
emission process
is analytically described as a quantum diffusion of the Bloch vector. 
It is shown that the evolution of the atomic wavefunction
can be controlled completely using the results of homodyne detection.
This allows the stabilization of a known quantum state or the creation of 
coherent states by a feedback mechanism. However, the
feedback mechanism can never compensate the dissipative effects of quantum
fluctuations even though the coherent state of the system is known at all
times. 

\end{abstract}

                    
\section{Introduction}

\label{sec:intro}

Homodyne detection is a procedure which allows the measurement of 
a quadrature component of the light field. In quantum optics, it has
long been applied to investigate squeezing effects of non-linear optical
systems. In that context it is not necessary to understand the 
time resolved properties of the homodyne detection process itself,
since the measurement is a time average over many photon emissions from
the light field source under investigation.
If homodyne detection is to be applied to the observation 
of individual quantum systems, however, a time resolved quantum mechanical
treatment of the projective measurements performed by a homodyne detection 
setup is necessary. Such a description not only helps to improve our
understanding of the quantum mechanics of photon emissions, but
also provides new methods of controlling the dynamics of quantum systems.

A complete quantum theory of homodyne detection must take into account both
the quantization of the field modes and the temporal evolution of the 
continuous light field entering the detector. The problem of 
quantization has been addressed in a number of publications 
\cite{Yue78,Bra90,Vog93,Lui96}. An approach to the problem of measuring a 
system as it 
evolves in time is presented by H. Carmichael in 
\cite{Car93}. It has been applied to a number of measurement scenarios, 
including homodyne and heterodyne detection \cite{Wis93,Wis96}, as well as
to feedback scenarios \cite{Wis94}. 

In  section \ref{sec:model} of this paper, we present an alternative 
derivation of this quantum trajectory
approach to time resolved observation of quantum systems by homodyne detection,
starting from the projective measurement postulate formulated in 
\cite{Yue78,Bra90,Vog93,Lui96}. 
In section \ref{sec:tsa} this measurement theory is then applied
to a description of the 
spontaneous emission process based on Wigner-Weisskopf theory, which takes
into account the field dynamics of the electromagnetic continuum of modes. 
This derivation provides a shortcut to the formulation of the back-action of 
the many
photon measurements on the two level atom and illustrates the problems 
of time resolution and of non-locality in quantum trajectories, which are 
discussed in \cite{Car93}. 
Once this control over the informatin loss induced by spontaneous emission
has been established, the observed changes in the atomic wavefunction can be 
compensated by feedback. This is investigated in section \ref{sec:cont}.
In section \ref{sec:concl}, the results are interpreted and conclusions are 
presented.

\section{A quantum mechanical description of time resolved homodyne 
detection}

\label{sec:model}  

\subsection{Time resolution and projective measurements in the
homodyne detection setup}

If we know that the linewidth of the emission we wish to observe is
much smaller than $1/\tau$, then we can conclude that a time resolution
of $\tau$ will be sufficient to observe the dynamics of photon emission
from the system. We therefore assume that the detectors used provide us
with reliable information about the photon numbers which entered each of
the detectors during a time interval $\tau$. In fact this is much closer 
to the experimental situation than the assumption of a continuous measurement
in which the arrival of a photon is detected with infinite time resolution.
For all time resolved measurements there is a time window of length $\tau$ 
which will usually not be shorter than a femtosecond. We can then describe the
homodyne detection as a projective measurement on a system of two light 
field modes of frequency $\omega_0$. The state to be measured is the 
product state of the input
field state $\mid \phi\rangle $ and the coherent state $\mid \alpha\rangle $ produced by the 
local oscillator. $\alpha$ is the complex field amplitude of the local
oscillator mode emitted during the time interval $\tau$. This amplitude is
related to the intensity $I$ emitted by the local oscillater via the 
relation $I=\alpha^*\alpha/\tau$. The measurement base is the photon number base 
of the
symmetric and anti-symmetric linear combinations of the input mode and
the local oscillator mode, $\mid n_+, n_-\rangle $. In terms of the input and local
oscillator fields, this state may be written as
\[
\mid n_+,n_-\rangle  = 
\]
\begin{equation}
\frac{1}{\sqrt{n_+!n_-!}} 
\left(\frac{\hat{a}^\dagger+\hat{b}^\dagger}{\sqrt{2}}\right)^{n_+}
\left(\frac{\hat{a}^\dagger-\hat{b}^\dagger}{\sqrt{2}}\right)^{n_-} \mid vacuum\rangle ,
\end{equation}
where $\hat{a}^\dagger$ is the creation operator of the local oscillator
light field mode ($\langle \alpha\mid \hat{a}^\dagger = \alpha^*\langle \alpha\mid $) and 
$\hat{b}^\dagger$ is the creation operator of the input field.

The effect of this measurement on the input field state $\mid \phi\rangle $ is a 
projective measurement on a set of states which is neither orthogonal nor normalized
within the subspace of interest.  
The state $\mid \mbox{P}(n_+,n_-)\rangle $ projected 
into by a measurement of $n_+$ and $n_-$ photons in the detectors is found by 
forming the scalar product with $\langle \alpha\mid $ in the subspace of the local 
oscillator field. Using the fact that $\langle \alpha\mid $ is an Eigenvector of 
$\hat{a}^\dagger$, one obtains 
\[
\mid \mbox{P}(n_+,n_-)\rangle  = 
\]
\begin{equation}
\frac{e^{-\mid \alpha\mid^2/2}}{\sqrt{n_+!n_-!}} 
\left(\frac{\alpha^*+\hat{b}^\dagger}{\sqrt{2}}\right)^{n_+}
\left(\frac{\alpha^*-\hat{b}^\dagger}{\sqrt{2}}\right)^{n_-} \mid vacuum\rangle .
\end{equation}
Note that even though this state is not normalized, the probability of a measurement of 
$n_+$ and $n_-$ is nevertheless given by
\begin{equation}
p(n_+,n_-) = \mid \langle \phi \mid \mbox{P}(n_+,n_-)\rangle \mid^2.
\end{equation}
The projection operators associated with these states form a complete probability
measure for the subsystem of the input light field.
It has been shown that the photon number difference $n_+-n_-$
corresponds very well to the quadrature component of the input field which 
is in phase with the local oscillator if the input field photon number is
much smaller than $\alpha^*\alpha$ (see for example \cite{Vog93}). Therefore, 
the classical interpretation of
a homodyne detection as a measurement of a quadrature component
of the input field also applies to the quantum mechanical measurement.  

\subsection{Application to low intensity light fields}

The light field sources we wish to investigate in the following   
will usually emit only zero or one photon ($n_{in}=0,1$) during the time 
interval $\tau$. 
If the input state is the vacuum state, the measurement probability is
given by
\begin{eqnarray}
\label{eq:gauss}
p_{vacuum}(n_+,n_-)&=&
\mid \langle vacuum\mid \mbox{P}(n_+,n_-)\rangle \mid^2 
\nonumber \\
&=& \frac{e^{-\mid \alpha\mid^2}}{n_+! n_-!}
\left(\frac{\alpha^*\alpha}{2}\right)^{(n_++n_-)}.
\end{eqnarray}
For a fixed sum of photons, $N = n_+ + n_-$, this probability distribution 
is the binomial distribution which results from the 
random scattering of the photons by the beam splitter. 
For large photon numbers $N \gg 1$, this probability distribution 
may be approximated by a Gaussian. The probability of measuring the 
photon number difference $\Delta n = n_+ - n_-$ regardless of the total photon number
is then given by
\begin{equation}
p_{vacuum}(\Delta n)=\frac{1}{\sqrt{2\pi\alpha^*\alpha}} 
\exp[-\frac{\Delta n^2}{2\alpha^*\alpha}].
\end{equation}
This probability distribution represents the vacuum fluctuations of
the in-phase quadrature of the incoming field. In the limit of very strong 
local oscillator 
fields, the measured value of this quadrature corresponding to the photon
number difference of $\Delta n$ is given by 
$\Delta n/2\mid \alpha\mid $, reproducing the expected  quantum uncertainty 
of $1/4$ in the quadrature component.

In the case of weak one photon contributions,
\begin{equation}
\mid \Phi_{\beta}\rangle  \approx \mid vacuum\rangle  + \beta \mid n_{in}=1\rangle ,
\end{equation}
the probability distribution is modified only slightly by the contributions
linear in $\beta$. 
To simplify the formalism,
the non-orthogonal measurement base of the homodyne detection
may be reduced to the components lying in the two dimensional 
subspace of zero or
one photon and the amplitude factor corresponding to the binomial distribution for
the vacuum can be approximated by a Gaussian. The resulting measurement base depends 
only on the photon
number difference $\Delta n$ and is given by
\[
\mid \mbox{P}(\Delta n)\rangle  = 
(2\pi\alpha^*\alpha)^{-1/4} \mid \mbox{P}(\Delta n)\rangle  
\]
\[
= 
(2\pi\alpha^*\alpha)^{-1/4} 
\exp[-\frac{\Delta n^2}{4\alpha^*\alpha}]
\]
\begin{equation}
\label{eq:Project}
\times
\left(\mid vacuum\rangle  + \frac{\Delta n}{\alpha^*}\mid n_{in}=1\rangle \right).
\end{equation}
Note that the weak field condition can be fulfilled for any input field
intensity by chosing a time scale $\tau$ which is much smaller than the 
average rate of photons corresponding to the intensity of the input field. 
Therefore, the simplified projection state $\mid \mbox{P}(\Delta n)\rangle $
can describe the time evolution of any homodyne detection scenario.
Applying equation (\ref{eq:Project}) to the weak
field state $\mid \Phi_{\beta}\rangle $ results in a probability distribution of
\begin{equation}
\label{eq:beta}
p_{\beta}(\Delta n)=\frac{1}{\sqrt{2\pi\alpha^*\alpha}} 
\exp[-\frac{(\Delta n-(\alpha^*\beta+\beta^*\alpha))^2}{2\alpha^*\alpha}].
\end{equation}
Effectively, the weak coherent field characterized by
$\beta$ shifts the Gaussian distribution by just the amount expected from
classical homodyne detection.


\section{Quantum diffusion of a two level atom}

\label{sec:tsa}
\subsection{Description of the emission process}

The coherent 
quantum dynamical evolution of an atomic system interacting with the
light field continuum is given by Wigner Weisskopf theory. The spatio-
temporal interpretation of this theory shows that emission processes 
can be described by a temporal evolution of the following type \cite{Hof95}:
\begin{mathletters}
\begin{equation}
\langle  E;vacuum\mid \Psi(t)\rangle  = e^{(-i\omega_0-\Gamma/2)t}
\end{equation}
\begin{equation}
\langle G;r\mid \Psi (t)\rangle  = \left\{ \begin{array}{c}-i
\sqrt{\frac{\Gamma}{c}}e^{(\Gamma/2+i\omega_0)(r/c-t)}
\quad\mbox{for}\quad 0 < r < ct\\
0\qquad \mbox{otherwise}\end{array}\right.
\end{equation}
\end{mathletters}
where $\mid E;vacuum\rangle $ is the state of the excited atom in the light 
field
vacuum and $\mid G;r\rangle $ is the state of the ground state atom with a 
photon
at a distance of $r$ from the atom. $\Gamma$ is the rate of spontaneous
emission. 

During a time interval $\tau$ which is much shorter than $1/\Gamma$,
the product state of the light field vacuum and an arbitrary linear 
combination of the excited state $\mid E\rangle $ 
and the ground state $\mid G\rangle $,
\begin{equation}
\mid \Psi(0)\rangle  = c_E \mid E;vacuum\rangle
                     + c_G \mid G;vacuum\rangle, 
\end{equation}
therefore evolves into an entangled 
state of the system and the light field, with the light field being 
in the vacuum state or in a single photon state. The single photon
wavefunction will be located within a distance of $c\tau$ from the atom.
Since the emission has a well defined frequency and angular dependence,
the mode into which the photon has been emitted is a well defined 
mode and the total state of atom and field can be written as
\begin{eqnarray}
\label{eq:photon}
\mid \Psi(\tau)\rangle  &=& 
c_E (1-(i\omega_0+\Gamma/2)\tau)\mid E;vacuum\rangle
\nonumber \\ & &
          + c_G \mid G;vacuum\rangle 
          + c_E \sqrt{\Gamma\tau}\mid G;n_0=1\rangle .
\end{eqnarray}
The amplitude factor of $\sqrt{\Gamma\tau}$ is found by normalizing the 
rectangular mode emitted by the system after a time interval of $\tau$.
This is done by dividing the amplitude of $\sqrt{\Gamma/c}$ emitted into
the spatially continuous one dimensional field by the 
amplitude of the normalized rectangular mode of length $c\tau$ which is 
equal to $\sqrt{1/c\tau}$.
Note that the amplitude of the emission depends on the square root of 
$\tau$, reflecting the linear increase of emission probability with time.

In homodyne detection, the frequency of the local oscillator is also 
$\omega_0$. Therefore, it is useful to transform to the interaction picture
using the time-dependent transformation
\begin{equation}
\label{eq:trans}
\mid \tilde{E}\rangle  = e^{-i\omega_0 t} \mid E\rangle .
\end{equation}
This effectively removes the terms oscillating with $\omega_0$ from
the system dynamics by describing the phase relation between the excited
state component and the ground state component of the system state 
not in terms
of an absolute phase but in terms of the phase relative to the local 
oscillator. Note that heterodyne detection may also be described by this 
formalism if the time scale $\tau$ is chosen so that the detuning 
$\delta\omega$ between the 
local oscillator and the system dynamics satisfies the
requirement that $\delta\omega \tau \ll 1$. The dynamical evolution of the 
system phase relative to the local oscillator phase can then be included in 
the measurement scenario.

The equation for the evolution of the wavefunction during
short time intervals $\tau$ is
\begin{eqnarray}
\label{eq:psi0}
\mid \Psi(\tau)\rangle  &=& c_E (1-\Gamma\tau/2)\mid \tilde{E};vacuum\rangle  
\nonumber \\
&& + c_G \mid G;vacuum\rangle 
   + c_E \sqrt{\Gamma\tau}\mid G;n_0=1\rangle .
\end{eqnarray}
The projection postulate of homodyne detection can now be applied
to the light field part of this correlated system-field state, resulting in 
an effective projection of the 
state of the atomic system into a state which depends on the measurement 
result of the homodyne detection.

Although the real physical measurement will take place many time intervals
$\tau$ later, after the light field wavefunction has traveled the distance to
the detectors, it is possible to interpret the measurement as 
an instantaneous projection on the local light field state, since the
light field signal, once emitted, will not interact with the atom again.
This artificial choice of the instant in which the state is projected
is consistent with the basic theory of quantum measurement, since it is not 
possible to distinguish between a projection at the time of a measurement and
a projection which anticipates the measurement. It is only our subjective 
expectation of causality which leads us to prefer placing the collapse after 
the measurement. 

Also the measurement described here is not continuous in the sense 
discussed in \cite{Wis93,Wis96}. The discrete measurements performed
are measurements of properties of the total interval $\tau$. For this 
reason $\tau$
is not written as $dt$, which would suggest the limit of
infinite time resolution. The experimental result corresponding to
the situation described here is a series of photon numbers without any
zero photon time intervals separating the time windows of each measurement.

If the wavefunction is regarded as an epistemological tool describing not
physical reality itself but only our knowledge of it, then 
the projective measurement at the atom is simply an expression of the 
information gained about the events at the atom, independent of the time at
which the information is actually obtained. In a completely relativistic
theory of measurement, this epistemological effect of future knowledge on
the state of the past must be considered, as was already pointed out by 
Einstein \cite{Ein31}. The arbitrary subdivision of the flow of time into
segments of duration $\tau$ is also an epistemological consequence as it
represents the time aspect of the space-time measurement base defined by
the experimental setup. Information about the atomic system dynamics on shorter
time scales is not obtained and cannot be included in the evolution of
the system wavefunction.

\subsection{Influence of the homodyne detection on the system dynamics}

For an arbitrary system with known initial wavefunction, the measurement 
protocol of repeated homodyne detections provides a complete description
of the evolution of the system wavefunction. In this sense, the method 
described here is a generalized quantum trajectory approach.
However, instead of using a master equation approach, Schroedinger´s
equation of the system-field interaction is solved and the 
homodyne detection events are described as projective measurements on
the correlated system-field wavefunction.
The detector simulated by this approach is not a single photon counting device
but rather corresponds to a photo diode with a high time-resolution.
Indeed, the measurement need not resolve single photon counts
to achieve a useful precision for predictions about the quantum system.

To investigate the dynamics of the atomic wavefunction observed by 
homodyne detection, it is useful to examine the effect of a single
projective measurement of photon number difference $\Delta n$ on an arbitrary
system wavefunction after an emission time segment of $\tau$. Before
the projective measurement, the correlated system-field wavefunction is
given by equation (\ref{eq:psi0}). 
After the projective measurement of $\Delta n$ the system is in a 
pure state described in the interaction picture by
\[
\mid \psi(\tau)\rangle  = \langle \mbox{P}(\Delta n) \mid \Psi (\tau)\rangle    
\]
\[
             = (2\pi\alpha^*\alpha)^{-1/4} 
             \exp[-\frac{\Delta n^2}{4\alpha^*\alpha}]
\]
\begin{equation} \times
          \left(c_E (1-\Gamma\tau/2)\mid \tilde{E}\rangle  
          +(c_G + c_E \sqrt{\Gamma\tau}\frac{\Delta n}{\alpha})\mid G\rangle \right).
\end{equation}
The squared length of this state vector is a measure of the probability 
of measuring $(n_+,n_-)$. Proceding as in the derivation of equation 
(\ref{eq:beta}) and assuming that $\Gamma\tau \ll 1$, it is possible to represent the deviation of this probability distribution from the vacuum state distribution by a Gaussian distribution
shifted by a term linear in $\sqrt{\Gamma\tau}$: 
\[
p(\Delta n)\approx
\]
\begin{equation}
\label{eq:prop}
 \frac{1}{\sqrt{2\pi\alpha^*\alpha}} 
             \exp[-\frac{(\Delta n- 
     \mid \alpha\mid \sqrt{\Gamma\tau}(c_Ec_G^*+c_E^*c_G))^2}{2\alpha^*\alpha}].
\end{equation}
Here and in the following, the phase relation between the local 
oscillator and the atomic dipole is defined so that it is zero if
both $c_E$ and $c_G$ are real and positive.
The small deviation from the quantum vacuum case are clearly related to the
dipole expectation value of the atomic system. Consequently, they indicate 
the dipole field emitted by the system.
The normalized change in the state of the system
is orthogonal to the initial state $\mid \psi (0)\rangle $ for small changes. It is 
found by projecting $\mid \psi(\tau)\rangle $  on the subspace orthogonal to 
$\mid \psi (0)\rangle $, normalizing by dividing by the amplitude of the parallel component.
\begin{equation}
\mid \delta\psi(\tau)\rangle  = 
\frac{\mid \psi(\tau)\rangle  -\mid \psi(0)\rangle \langle \psi(0)\mid \psi(\tau)\rangle  }
     {\langle \psi(0)\mid \psi(\tau)\rangle }
\end{equation}
The normalized expression for the system wavefunction is then given by
$\mid \psi(0)\rangle  + \mid \delta \psi(\tau)\rangle $. Applying the 
condition that $\Gamma\tau \ll 1$ we neglect all terms above second order in 
$\sqrt{\Gamma\tau}$. The change of the wavefunction $\mid \delta\psi (\tau)
\rangle$ within the time interval $\tau$ conditioned by the measurement of
$\Delta n$ is then given by
\begin{eqnarray}
\label{eq:svdiff}
\mid \delta\psi(\tau)\rangle 
                   &=& -\sqrt{\Gamma\tau}\frac{\Delta n}{\mid\alpha\mid}
c_E^2\left(c_G^* \mid \tilde{E}\rangle  - c_E^* \mid G\rangle \right)\nonumber \\
                   & & + \Gamma\tau \frac{\Delta n^2}{\alpha^*\alpha }
c_E^3c_G^*\left(c_G^* \mid \tilde{E}\rangle  - c_E^* \mid G\rangle \right)\nonumber \\
                   & & -\frac{\Gamma\tau}{2}
c_E  c_G  \left(c_G^* \mid \tilde{E}\rangle  - c_E^* \mid G\rangle \right).
\end{eqnarray}

The dominant effect within one time interval is given by the diffusion term
proportional to $\sqrt{\Gamma\tau}\Delta n$. The average of these diffusion
steps is only proportional to $\Gamma\tau$ however, since the expectation 
value of $\Delta n/\alpha$ is of the order of $\sqrt{\Gamma\tau}$. 
On the measurement timescale $\tau$, the quantum fluctuations therefore 
cause a random walk type diffusion of the atomic state, just as one would 
expect from classical noise. However, the length and the phase of the
diffusion step are functions of the initial system state.
On timescales of $1/\Gamma$, the non-zero average of $\Delta n/\alpha$,
the diffusion term proportional to the square of the measurement, and 
the deterministic drift terms contribute to the change in the system state.
All of these terms give rise to a slow drift of the wavefunction towards the 
ground state.
To illustrate the dependence of the diffusion constant on the state of the 
atomic system and to identify the drift terms
we will now formulate this dependence of these processes 
in terms of the Bloch vector.   

\subsection{Quantum diffusion of the Bloch vector}

The dynamics of two level systems can be visualized using the Bloch
vector representation. This three dimensional vector incorporates
the excitation (population inversion), the dipole and the dipole current 
of the two level atom
as its orthogonal components. When the time dependent transformation
given in equation (\ref{eq:trans}) is used, the two components orthogonal
to the excitation $s_z$ describe the in-phase and the ($\pi/2$) out-of-phase 
components, respectively, of the dipole oscillations relative to the local 
oscillator.
\begin{mathletters}
\begin{eqnarray}
s_{x} &=& 2 Re(\langle \psi\mid \tilde{E}\rangle \langle G\mid \psi\rangle ) \\
s_{y} &=& 2 Im(\langle \psi\mid \tilde{E}\rangle \langle G\mid \psi\rangle ) \\
s_{z} &=& \mid \langle \tilde{E}\mid \psi\rangle \mid^2 - \mid \langle G\mid \psi\rangle \mid^2
\end{eqnarray}
\end{mathletters}
If $\alpha$ is a real number, $s_{x}$ is the in-phase component of the
atomic dipole and $s_{y}$ is the out-of-phase component.
The diffusion step associated with a measurement result of $\Delta n$ in
terms of the Bloch vector is derived from $\mid \psi(0)\rangle $ and $\mid \delta\psi(\tau)\rangle $
by using
\begin{mathletters}
\begin{eqnarray}
\delta s_{x} &=& 2 Re(\langle \psi(0)\mid \tilde{E}\rangle \langle G\mid \delta\psi(\tau)\rangle \nonumber \\ &&
              +\langle \delta\psi(\tau)\mid \tilde{E}\rangle \langle G\mid \psi(0)\rangle ) \\
\delta s_{y} &=& 2 Im(\langle \psi(0)\mid \tilde{E}\rangle \langle G\mid \delta\psi(\tau)\rangle  \nonumber \\ &&
              +\langle \delta\psi(\tau)\mid \tilde{E}\rangle \langle G\mid \psi(0)\rangle ) \\
\delta s_{z} &=& 2 Re(\langle \psi(0)\mid \tilde{E}\rangle \langle \tilde{E}\mid \delta\psi(\tau)\rangle  \nonumber \\ &&
              -\langle \psi(0)\mid G\rangle \langle G\mid \delta\psi(\tau)\rangle ).
\end{eqnarray}
\end{mathletters}

The change in the Bloch vector $\delta {\bf s}$ can be expressed in terms
of the Bloch vector ${\bf s}$ corresponding to $\mid \psi(0)\rangle $. 
The result is the diffusion step of the Bloch vector,
\begin{eqnarray}
\label{eq:diffstep}
\left(\begin{array}{c}\delta s_{x}\\ \delta s_{y}\\ \delta s_{z}\end{array}
\right) &=&  \sqrt{\Gamma\tau}\frac{\Delta n}{\mid\alpha\mid}
\left(\begin{array}{c}1+s_z-s_x^2\\-s_xs_y\\-s_x(1+s_z)\end{array}
\right) \nonumber \\ && 
            -\Gamma\tau \frac{\Delta n^2}{\alpha^*\alpha} s_x
\left(\begin{array}{c}1+s_z-s_x^2\\-s_xs_y\\-s_x(1+s_z)\end{array}
\right) 
\nonumber \\
        & & +\frac{\Gamma\tau}{2}\frac{\Delta n^2}{\alpha^*\alpha} (1+s_z)
\left(\begin{array}{c}s_xs_z\\s_ys_z\\s_z^2-1\end{array}
\right)  \nonumber \\ &&
            +\frac{\Gamma\tau}{2} 
\left(\begin{array}{c}s_xs_z\\s_ys_z\\s_z^2-1\end{array}
\right) 
. 
\end{eqnarray}
As in the case of the state vector representation, the Bloch vector 
diffusion step is composed of a fast random walk diffusion on the
time scale of $\tau$ and drift terms on a time scale of $1/\Gamma$.
What is more apparent in the Bloch vector representation is the possible
separation into phase sensitive and phase independent changes. 

\subsection{Interpretation of the contributions to the diffusion step}

To interpret the diffusion step, it is useful to analyze the seperate
contributions in more detail. 
In particular it is helpful to investigate
the diffusion step component representing the random walk caused by
the fluctuating light field,
\begin{equation}
\left(\begin{array}{c}\delta s_{x}\\ \delta s_{y}\\ \delta s_{z}\end{array}
\right)_{fluctuation} =  \sqrt{\Gamma\tau}\frac{\Delta n}{\mid\alpha\mid}
\left(\begin{array}{c}1+s_z-s_x^2\\-s_xs_y\\-s_x(1+s_z)\end{array}
\right). 
\end{equation}
Figure \ref{diffusion} illustrates this diffusion on the Bloch sphere.
In order to analyse the diffusive motion of the Bloch vector, one may 
separate the absolute value from the direction of the diffusion. 
The diffusion constant is then given by
\begin{equation}
\frac{\langle \delta{\bf s}^2\rangle }{\tau}= \Gamma \left(1+s_z\right)^2.
\end{equation}
The magnitude of the diffusion is therefore independent of the phase
relation between the local oscillator and the atomic dipole. 
It does depend on the excitation of the atom however.
It is maximal for the excited state and zero for the ground state, indicating
that the ground state will not interact with the field vacuum.
The direction of $\delta{\bf s}$ is given by
\begin{mathletters}
\begin{eqnarray}
\frac{\delta {\bf s}}{\mid \delta {\bf s}\mid } &=& \frac{1}{1+s_z}\left(\begin{array}{c}
1+s_z-s_x^2 \\ - s_xs_y \\ -s_x(1+s_z) \end{array}\right)
\nonumber \\ 
&=& \frac{s_y}{\sqrt{s_x^2+s_y^2}}\;{\bf \hat{e}_\perp} + \frac{s_x}
{\sqrt{s_x^2+s_y^2}}
\;{\bf \hat{e}_\parallel},
\end{eqnarray}
where ${\bf \hat{e}_\perp}$ is the unit vector perpendicular to both 
${\bf s}$ and
the z-axis and ${\bf \hat{e}_\parallel}$ is the unit vector perpendicular 
to both
${\bf s}$ and ${\bf \hat{e}_\perp}$.
\begin{eqnarray}
{\bf \hat{e}_\perp}&=&\frac{1}{\sqrt{s_x^2+s_y^2}}\left(\begin{array}{c}
s_y \\ -s_x \\ 0 \end{array}\right) 
\nonumber \\ &&\mbox{and} \nonumber \\ 
 {\bf \hat{e}_\parallel}&=&\frac{1}{\sqrt{s_x^2+s_y^2}}\left(\begin{array}{c}
s_xs_z \\ s_ys_z \\ -s_x^2-s_y^2 \end{array}\right).
\end{eqnarray}
\end{mathletters}
As can be seen from these results, a Bloch vector which is in phase
with the local oscillator will show random rotations in the $s_x,
s_z$ plane. A Bloch vector polarized out of phase
with the local oscillator by $\pm \pi/2$ ($s_x=0$) will undergo phase
diffusion only, with $s_z$ remaining constant. The ratio between the
diffusion constant of pure phase diffusion and the diffusion constant 
of excitation diffusion is given by $(s_y/s_x)^2 = \tan(\phi)^2$, where
$\phi$ is the phase difference between the local oscillator and the dipole
oscillations of the atom.
Note that a randomly varying classical in phase field would give rise to
Rabi oscillations around $s_y$ with a step length proportional to 
$\sqrt{1-s_x^2}$. The properties that only phase diffusion occurs for
$s_x=0$ and that the rotations in the $s_y=0$ plane preserve phase are
also properties of Rabi rotations induced by classical fields. However,
the step length dependence of $1+s_z$ clearly indicates a difference 
between the effects of quantum noise and of classical noise.  

On a timescale of $1/\Gamma$, the contribution of the random walk is
given by the non-vanishing expectation value of the homodyne detection
photon number difference. The probability distribution 
of the measurement
results is approximately given by a Gaussian with 
\begin{mathletters}
\begin{equation}
\langle \frac{\Delta n}{\mid\alpha\mid}\rangle  = \sqrt{\Gamma\tau}s_x
\end{equation}
\begin{equation}
\langle \frac{\Delta n^2}{\alpha^*\alpha}\rangle  = 1.
\end{equation}
\end{mathletters}
On a timescale of $1/\Gamma$ a  large number of measurements will have 
been performed since $\tau \ll 1/\Gamma$. Therefore, the net change of the
system state can be evaluated using these expectation values. 
The random walk drift is then given by
\begin{equation}
\left(\begin{array}{c}\langle \delta s_{x}\rangle \\
\langle \delta s_{y}\rangle \\
\langle \delta s_{z}\rangle \end{array}
\right)_{fluctuation}  = \Gamma\tau s_x
\left(\begin{array}{c}1+s_z-s_x^2\\-s_xs_y\\-s_x(1+s_z)\end{array}
\right). 
\end{equation}
This is exactly compensated by the other phase sensitive contribution 
effective on a time scale of $1/\Gamma$,
\begin{equation}
\left(\begin{array}{c}\langle \delta s_{x}\rangle \\
\langle \delta s_{y}\rangle \\
\langle \delta s_{z}\rangle \end{array}
\right)_{compensation}  = -\Gamma\tau \langle \frac{\Delta n^2}
{\alpha^*\alpha}\rangle s_x
\left(\begin{array}{c}1+s_z-s_x^2\\-s_xs_y\\-s_x(1+s_z)\end{array}
\right). 
\end{equation}
This compensation has a clear physical reason. The dynamics of the system
should only depend on the incoming field. However, the measurement is 
being performed on the field coming from the atomic system which is a
sum of the fields passing the atom and the dipole field emitted by the atom.
Although it is impossible to seperate the contributions in a quantum
mechanical measurement, the long term averages
can actually be corrected by subtracting the averaged dipole field 
contribution from the measured fields. In effect, this leaves only two
drift terms which influence the $1/\Gamma$ time scale dynamics.

The drift term which is not dependent on the measurement result $\Delta n$
is given by
\begin{equation}
\left(\begin{array}{c}\delta s_{x}\\ \delta s_{y}\\ \delta s_{z}\end{array}
\right)_{dipole}  = \frac{\Gamma\tau}{2} 
\left(\begin{array}{c}s_zs_x\\s_zs_y\\-s_x^2-s_y^2\end{array}
\right). 
\end{equation}
This term describes a reduction of $s_z$ as a function of $s_x^2+s_y^2$.
Since $s_z$ is the energy expectation value of the atomic system and
$s_x^2+s_y^2$ is the square of the atomic dipole, this process describes 
a loss of energy corresponding to the emission of radiation from a 
classical oscillating dipole.
It is therefore equivalent to the equations describing super radiance
\cite{Dic54,Bon71}, as
well as to a semiclassical textbook approach to spontaneous emission
\cite{Ger86}. 
Unlike the other terms, this term is entirely free of quantum fluctuation
effects. It is completely deterministic and depends only on the expectation
values of the atomic system state. It is a fascinating feature of the 
quantum theoretical formulation of spontaneous emission applied in this
investigation that it automatically produces such a semiclassical term from
the simple photon emission described by equation (\ref{eq:photon}).

The three terms discussed up to this point can be understood in terms of
the action of the incoming quantum fluctuations on the system and in terms 
of the semiclassical emission caused by the dipole oscillations. The
fourth term is more difficult to understand since it depends on the quantum
fluctuation measurements but is not sensitive to the phase of the local 
oscillator. It seems to be a mixed effect of dipole emission and quantum 
fluctuations, possibly related to the quantum fluctuations of the atomic
dipole. The contribution of this mixed term to the dynamics is given by

\begin{equation}
\left(\begin{array}{c}\langle \delta s_{x}\rangle \\
\langle \delta s_{y}\rangle \\
\langle \delta s_{z}\rangle \end{array}
\right)_{mixed}  = \frac{\Gamma\tau}{2} \langle \frac{\Delta n^2}{\alpha^*\alpha}\rangle  (1+s_z)
\left(\begin{array}{c}s_zs_x\\s_zs_y\\-s_x^2-s_y^2\end{array}
\right). 
\end{equation}
The sum of the drift dynamics to be expected on a timescale of $1/\Gamma$ is
given by the sum of the dipole contribution and the mixed term. The total
drift vectors resulting from this sum are shown in figure \ref{drift}.

\subsection{Contributions to the exponential decay of the average excitation}

At first it may seem confusing that a simple spontaneous emission from an
isolated atom should give rise to such a complicated variety of dynamical 
effects. After all, the Bloch vector dynamics of an ensemble of atoms is 
described by
\begin{equation}
\frac{d}{dt}
\left(\begin{array}{c}\langle s_{x}\rangle \\
\langle s_{y}\rangle \\
\langle s_{z}\rangle \end{array}\right)
= - \Gamma  \left(\begin{array}{c}\langle s_x\rangle /2\\
\langle s_y\rangle /2\\
\langle s_z\rangle +1\end{array}\right). 
\end{equation}
If one separates the time derivative of the Bloch vector into a component
orthogonal to the Bloch vector and one parallel to it, however, the
connection to the dynamics observed in the homodyne detection becomes 
apparent: 
\begin{eqnarray}
\frac{d}{dt}
\left(\begin{array}{c}\langle s_{x}\rangle \\
\langle s_{y}\rangle \\
\langle s_{z}\rangle \end{array}\right)
&=&   \frac{\Gamma}{2}\frac{(2+\langle s_z\rangle )}{\langle {\bf s}\rangle ^2}
    \left(\begin{array}{c}\langle s_z\rangle \langle s_x\rangle \\
\langle s_z\rangle \langle s_y\rangle 
                        \\
\langle s_z\rangle ^2-\langle {\bf s}\rangle ^2\end{array}\right)\nonumber \\
& & - \frac{\Gamma}{2}\frac{(\mid \langle {\bf s}\rangle \mid +\langle s_z\rangle )^2}{\langle {\bf s}\rangle ^2}
    \left(\begin{array}{c}\langle s_x\rangle \\
\langle s_y\rangle \\
\langle s_z\rangle \end{array}\right). 
\end{eqnarray}
For $\mid \langle {\bf s}\rangle \mid =1$, the expected reduction of the length of the Bloch
vector is an effect of the diffusion caused by the 
${\bf \delta s}_{fluctuations}$ term. Its rate is given by one half of
the diffusion constant $\Gamma \left(1+s_z\right)^2$.
The rotation of the Bloch vector is given by the sum of the two drift terms,
${\bf \delta s}_{dipole}$ and ${\bf \delta s}_{mixed}$. 
For $\mid \langle {\bf s}\rangle \mid =1$, the average expected change
in $s_z$ can be separated into contributions related to the diffusion of the 
Bloch vector, to the dipole emission, and to the mixed term,
\begin{mathletters}
\begin{equation}
\frac{d}{dt}\langle s_z\rangle  = -\Gamma(\langle s_z\rangle +1) 
(q_{diffusion} + q_{dipole} + q_{mixed})
\end{equation}
\begin{eqnarray}
q_{diffusion} &=& \frac{1}{2}(1+s_z)s_z \\
q_{dipole} &=& \frac{1}{2}(1-s_z) \\
q_{mixed} &=& \frac{1}{2}(1-s_z^2). 
\end{eqnarray}
\end{mathletters}
Figure \ref{contributions} shows this change in the relative contributions
to the exponential decay of the atomic excitation as a function of $s_z$.
For the excited state, $s_z=+1$, the exponential
relaxation of $s_z$ is a result of the diffusion of the Bloch vector due
to quantum fluctuations, while for states 
close to the ground
state, $s_z\approx-1$, the exponential relaxation to $s_z=-1$ is dominated
by the semiclassical dipole emission. For maximally polarized states, 
$s_z=0$, the emission contributions of the dipole term and of the mixed term
are equally strong, while the diffusion has no effect. Note that for
$-1<s_z<0$, the diffusion caused by the quantum fluctuations actually tends 
to excite the atom. However, this
effect is compensated by the mixed terms and the dipole emission.


\section{Compensating quantum fluctuations by feedback}

\label{sec:cont}

\subsection{Feedback setup}

Section \ref{sec:tsa} describes the measurement of quantum fluctuations 
propagating away from an atomic two level system by homodyne detection
and the implications of the measurement for the dynamics of the atomic state.
Since maximal knowledge is obtained about the field state, the atomic system
state is in a pure state after the measurement. This suggests the possibility
of using feedback based on the measurement results as a means to manipulate 
the quantum state. A general formalism for this procedure based on the master
equation approach to quantum trajectories has been presented in \cite{Wis94} and
was applied to light fields in an open cavity in \cite{Wis93a}. In the following, we will discuss the possibility and the physical problems involved 
in applying a feedback scheme to the two level atom.
 
On a timescale of 
$\tau$, the homodyne detection does not give any information
about the state of the atom. $p(\Delta n)$ is nearly equal to the vacuum 
probability distribution. The information obtained in a single field 
measurement
is therefore not information about the atomic system itself. Instead, it
is information about the quantum fluctuations acting on the system.
This sensitivity to the vacuum fluctuations (as opposed to the system state)
is ideal for stabilizing quantum states by reversing the effects of the 
measurement process \cite{Ued92}.
If we were dealing with a classical system it would be possible to
measure the exact forces involved. A feedback mechanism could then compensate 
the forces, thereby removing the effects of the incoming noise. In quantum 
mechanics however, the quadratures of the light field do not commute. 
Therefore the force acting on the atomic system can never be known 
completely. In the balanced homodyne detection setup discussed here we only 
know the quadrature of the light field which is in phase with the local 
oscillator. In a classical system, this lack of information
would make control impossible. In the quantum system however, we can still
achieve perfect control of the atomic wavefunction.    

In order to manipulate the known atomic state the electromagnetic field 
at the atom may be modified. For example, a resonant coherent driving field 
may be
coupled to the atomic system. In order to ensure that this field has a fixed
phase relation with the local oscillator used in the measurement it would
be natural to utilize the same coherent light field source in the homodyne
detection setup and in the feedback loop. Since the local oscillator field is 
very strong, only a negligibly small portion needs to be redirected in order 
to achieve control of the system dynamics. In fact, since the strength of
the field expectation value needs to compensate for quantum fluctuations only,
the portion of the local oscillator intensity used for this purpose is 
approximately given by
$1/(4 \alpha^*\alpha)$. Note that it is possible to add the effect
of the Rabi rotations caused by the feedback field and the effects of 
the quantum fluctuations because the Heisenberg equations of motion for the
fully quantum mechanical field-atom interaction are linear in the field 
variables. The dynamical effects of the feedback and of the diffusive
evolution are therefore seperable.

A controllable reflector can be used to coherently manipulate the system
depending on the measurement results of the homodyne detection. This feedback
modifies the dissipative dynamics of the atomic system. If the delay between
emission and feedback is much shorter than $1/\Gamma$, it is possible to 
compensate the effects of fluctuations on a known system state. 
Note that this either requires a low decay rate $\Gamma$ or a very fast 
feedback loop. In an optical setup a typical unmodified 
lifetime of nanoseconds would require a feedback loop  much 
shorter than 10 cm in length, so the light field signal can return in time,
with the purely optical dissipative photon detectors integrated into this
loop. 

The effect of a delay time of $\Delta t = m \tau$ between the
emission and the arrival of the corresponding feedback signal at the atom
can be estimated by considering
the number of uncompensated diffusion steps $m$ which occur during the delay
time. The probability distribution over the sum of the measurement results 
$\delta n$ accumulated within that 
uncontrolled interval is given by a Gaussian with a standard deviation of
$\sqrt{m}\mid\alpha\mid$. This may be multiplied with the expected 
uncompensated diffusion step length at the stabilized point on the Bloch
sphere in order to determine the reliability of feedback stabilization with
a non vanishing delay time. In the following, however, we shall concentrate
on the description of effectively instantaneous feedback.

Because the total light field propagating away from the atomic system is
measured in the homodyne detection, the reflected control field will also 
be measured in the subsequent time interval. This must be subtracted from 
the measurement signal for further feedback, since the feedback effect itself 
should not be compensated. Note that the linearity of the field dynamics
involved actually allows this seperation despite the fact that the feedback
field and the quantum fluctuations interact with the system in a qualitatively
different manner as described below. Intuitively, one would expect 
a complete compensation of the fluctuations if the averaged measurements
of $\Delta n$ are zero. However, we will see that this is not so, owing
to the fact that only one quadrature of the quantum fluctuations can
be compensated.

\subsection{Effects of feedback on timescales shorter than $1/\Gamma$}

Although a number of feedback scenarios may be discussed even in the simple
context of the two level atom, we will now concentrate on the possibility of
stabilizing a known state with $s_y = 0$ by effectively instantaneous feedback.
The feedback is given by a coherent driving field inducing Rabi rotations
around the $s_y$ axis. The effect of this feedback experienced by
the system during the time interval $\tau$ can be written as
\begin{equation}
\left(\begin{array}{c}\delta s_{x}\\ \delta s_{y}\\ \delta s_{z}\end{array}
\right)_{feedback} =  2\sqrt{\Gamma\tau} f(\Delta n)
\left(\begin{array}{c} s_z\\ 0 \\-s_x \end{array}
\right), 
\end{equation}
where $f(\Delta n)$ is the feedback field describing the response of the
coherent control to the most recent measurement result. $\Delta n$ is the 
measurement result obtained from the quantum 
fluctuations acting on the system in the previous time interval. Since the
time intervals are small on the scale of the system dynamics, however, we
will simplify matters by summing the effects of the diffusion step and its
subsequent feedback as if they occurred in the same time interval to obtain
the effective total diffusion.

The homodyne detection measurement in the following time interval will be 
modified
by the feedback field. The change in the $\Delta n$
measurement caused by the feedback in the next measurement, $\delta_{next}$,
may be determined using equation(\ref{eq:beta}):
\begin{equation}
\delta_{next} = 2 \mid \alpha\mid  f(\Delta n).
\end{equation}

In order to stabilize a given system state ${\bf s}$, $f(\Delta n)$ must be
chosen to compensate the diffusion step, i.e.
\begin{equation}
\left(\begin{array}{c}\delta s_{x}\\ \delta s_{y}\\ \delta s_{z}\end{array}
\right)_{fluctuation} 
+ \left(\begin{array}{c}\delta s_{x}\\ \delta s_{y}\\ \delta s_{z}\end{array}
\right)_{feedback}=0. 
\end{equation}
For $s_y = 0$, this condition is fullfilled if
\begin{equation}
f(\Delta n)=-(1+\bar{s}_z)\frac{\Delta n}{2\mid \alpha\mid },
\end{equation}
where $\bar{s}_z$ is the Bloch vector component of the stabilized state. Note
that $f(\Delta n)$ is not a function of the present state of the atomic system.

A particularily interesting case is obtained when the feedback is chosen
to compensate for the maximal fluctuation effects possible, stabilizing
the excited state of the two level atom. If the system is not in the 
excited state, the diffusion step associated with a homodyne detection 
event of $\Delta n$ is now given by the sum of the original diffusion step 
and the feedback,
\begin{eqnarray}
\left(\begin{array}{c}\delta s_{x}\\ \delta s_{y}\\ \delta s_{z}\end{array}
\right)_{eff.} &=&  \sqrt{\Gamma\tau}\frac{\Delta n}{\mid \alpha\mid }
\nonumber \\[1em] &&
\times
(\left(\begin{array}{c}1+s_z-s_x^2\\-s_xs_y\\-s_x(1+s_z)\end{array}
\right)+\left(\begin{array}{c}-2s_z\\ 0 \\ 2s_x\end{array}
\right))\nonumber \\[1em]
&=& \sqrt{\Gamma\tau}\frac{\Delta n}{\mid \alpha\mid }
\left(\begin{array}{c}1-s_z-s_x^2\\-s_xs_y\\ s_x(1-s_z)\end{array}
\right). 
\end{eqnarray}
As shown in figure \ref{fbdiff1}, this effective diffusion step is the exact 
inversion of the diffusion
step without feedback. The roles of $s_z=-1$ and of $s_z=+1$ have been 
reversed. Consequently,
the sensitivity of the system to quantum fluctuations is now maximal at
the ground state, $s_z=-1$. The diffusion constant is proportional to
$(1-s_z)^2$.
The in-phase quantum fluctuations have been over-compensated in this case:
\begin{equation}
\Delta n+\delta_{next}=-\Delta n.
\end{equation}
This indicates, that the reversal of the diffusion effect has been achieved by
answering the fluctuations of the in-phase quadrature with an opposite field
of double strength, effectively reversing the sign of that field component, 
while the
unknown fluctuations in the out-of-phase quadrature are unmodified.

Another interesting scenario is the stabilization of $s_z=0$, because it 
corresponds to the simple minded compensation of quantum fluctuations by
chosing a feedback field with the negative amplitude of the quadrature 
measured in the homodyne detection. 
The effective diffusion step now reads
\begin{equation}
\left(\begin{array}{c}\delta s_{x}\\ \delta s_{y}\\ \delta s_{z}\end{array}
\right)_{eff.} =  \sqrt{\Gamma\tau}\frac{\Delta n}{\mid \alpha\mid }
\left(\begin{array}{c}1-s_x^2\\-s_xs_y\\ -s_xs_z\end{array}
\right). 
\end{equation}
As shown in figure \ref{fbdiff0}, this diffusion law has rotational symmetry 
around the $s_x$ axis. The ratio
of $s_y/s_z$ is a constant. The diffusion is always directed towards one of
the two stabilized poles with $s_x=\pm 1$. The absolute value of the 
diffusion constant for the stabilization of $s_z=0$ is thus given by

\begin{equation}
\frac{\langle \delta{\bf s}^2\rangle }{\tau}= \Gamma(1-s_x^2).
\end{equation}

In the general case of a stabilization of $\bar{s}_z=\cos \theta$, where
$\theta$ is the angle between the stabilized Bloch vector and the $s_z$ axis,
the diffusion step is
\begin{eqnarray}
\left(\begin{array}{c}\delta s_{x}\\ \delta s_{y}\\ \delta s_{z}\end{array}
\right)_{eff.} &=&  \sqrt{\Gamma\tau}\frac{\Delta n}{\mid \alpha\mid }
(s_{y}\left(\begin{array}{c}s_y\\-s_x\\0\end{array}
\right)
\nonumber \\ &&
+(s_{z}-\cos \theta)\left(\begin{array}{c} s_z\\ 0 \\ -s_x\end{array}
\right))
\end{eqnarray}
and the diffusion constant is given by 
\begin{equation}
\frac{\langle \delta{\bf s}^2\rangle }{\tau}=\Gamma(1-s_x^2-2\cos \theta s_z + 
\cos^2\theta (s_x^2+s_z^2)).
\end{equation}
By varying the feedback it is therefore possible to suppress the diffusive 
dynamics for an arbitrary state in the $s_x,s_z$ plane.

\subsection{Effects of the drift terms on timescales of $1/\Gamma$}

Although the feedback described above can completely suppress the random walk
dynamics induced by the quantum fluctuations on a timescale of $\tau$ for an
arbitrary system state, it is necessary to consider the effects of the drift
terms if stability on longer time scales is to be obtained as well. The 
complete diffusion step, including a feedback field of 
$f(\Delta n) = -(1+\cos\theta)\Delta n/2\mid \alpha \mid $,
is given by
\begin{equation}
\left(\begin{array}{c}\delta s_{x}\\ \delta s_{y}\\ \delta s_{z}\end{array}
\right) =
  \left(\begin{array}{c}\delta s_{x}\\ \delta s_{y}\\ \delta s_{z}\end{array}
\right)_0
- \sqrt{\Gamma\tau}\frac{\Delta n}{\mid \alpha \mid}(1+\cos\theta)
\left(\begin{array}{c}s_z\\0\\-s_x\end{array}
\right), 
\end{equation}
where $\delta {\bf s}_0$ denotes the diffusion step without feedback, as
given by equation (\ref{eq:diffstep}). The drift is consequently modified
by the non-zero average of $\Delta n$ in the feedback term.
The resulting drift is the sum of the drift without feedback and the
feedback drift term. It may be written as the sum of a rotation around 
the $s_x$ axis and a rotation around the $s_y$ axis,
\begin{eqnarray}
\left(\begin{array}{c}\langle \delta s_{x}\rangle \\
\langle \delta s_{y}\rangle \\
\langle \delta s_{z}\rangle \end{array}
\right) &=&
  \frac{\Gamma\tau}{2}(2+s_z)s_y
\left(\begin{array}{c}0\\s_z\\-s_y\end{array}\right)
 \nonumber \\ &&
+ \frac{\Gamma\tau}{2}(s_z-2\cos\theta)s_x
\left(\begin{array}{c}s_z\\0\\-s_x\end{array}\right). 
\end{eqnarray}
For the stabilized state with $\bar{s}_x=\sin\theta, \bar{s}_y=0,
\bar{s}_z=\cos\theta$,
the effective drift is equivalent to a Rabi rotation around the $s_y$
axis of $(\Gamma\tau/2)\sin \theta \cos \theta$ per time interval $\tau$. 
This effect can be 
compensated by a constant driving field. The field strength $f_0$ necessary
for this purpose is given by
\begin{equation}
f_0 = \frac{\sqrt{\Gamma\tau}}{4}\sin\theta\cos\theta.
\end{equation}
This driving field ensures that the stabilized state is a stationary solution
of the drift-diffusion dynamics generated by the feedback setup.
However, the drift terms may also
cause problems if they amplify small deviations from this stabilized state.

The stability analysis near $s_x=\sin\theta, s_y=0,s_z=\cos\theta$ can be
performed by linearizing the drift dynamics of small deviations. The deviation
in the $s_x,s_z$ plane is appropriately described by the angular variable
$\epsilon$, such that $s_z = \cos (\theta + \epsilon)$ and 
$s_x = \sin (\theta + \epsilon)$. The linear stability analysis then 
shows that for small $\epsilon$, the drift $<\delta \epsilon>$ per time 
interval $\tau$ is
\begin{equation}
\langle \delta\epsilon\rangle  = - \frac{\Gamma\tau}{2}\epsilon .
\end{equation}
Therefore the drift terms always stabilize the state with suppressed 
fluctuations against rotations in the $s_x,s_z$ plane. In the $s_y,s_z$
plane, the situation is different however. The linearized drift dynamics 
of small $s_y$ is described by
\begin{equation}
\langle \delta s_y\rangle  = \frac{\Gamma\tau}{2}(2+\cos\theta)\cos\theta s_y.
\end{equation}
This indicates stability for all $\cos\theta < 0$, i.e. any state which has
a negative $s_z$ component. If states of higher excitation are to be 
stabilized, any small deviation from $s_y=0$ in the initial preparation is
amplified exponentially on a timescale of $1/\Gamma$. This indicates that
long term stability of such inverted states cannot be achieved. 
 
An interesting aspect of the quantum trajectories associated with the
measurement protocols obtained from the homodyne detection is revealed
in this critical problem of quantum control. If the initial state is only
known within an error margin, the trajectories may diverge and amplify this 
error margin even though the trajectories are fully deterministic functions
of the initial state. Apparently, there is an aspect of deterministic chaos 
in this small quantum system when its dynamics is analyzed using homodyne
detection.
It seems that the fully polarized states, $s_x=\pm1, 
s_y=s_z=0$, are ideal for stabilization, since they can be observed in the 
measurement results. If the system drifts away from its stabilized state,
this can be observed in the homodyne detection and may be corrected, either
by applying static fields to shift the phase of the atomic system or by 
shifting the phase of the local oscillator. In the following we will
investigate the example of feedback stabilization for these maximally coherent
states.

\subsection{Stabilization of the dipole eigenstates}

For $\cos\theta = 0$, the stabilized states are $s_x=\pm 1$, the 
eigenstates of the dipole component oscillating in phase with the local
oscillator. Also, this case is special because it corresponds to the classical
idea of feedback compensation: $\Delta n + \delta_{next}= 0$. At the same
time, the average of $\Delta n$ is a measure of $s_x$, indicating both the
sign of the Bloch vector component stabilized and the success or 
failure of the stabilization attempt.
The modified total diffusion step including feedback is given by
\begin{eqnarray}
\left(\begin{array}{c}\delta s_{x}\\ \delta s_{y}\\ \delta s_{z}\end{array}
\right) &=&  \sqrt{\Gamma\tau}\frac{\Delta n}{\mid\alpha\mid}
\left(\begin{array}{c}1-s_x^2\\-s_xs_y\\-s_xs_z\end{array}
\right)  \nonumber \\ &&
            -\Gamma\tau \frac{\Delta n^2}{\alpha^*\alpha} s_x
\left(\begin{array}{c}1-s_x^2\\-s_xs_y\\-s_xs_z\end{array}
\right) 
\nonumber \\
        & & -\Gamma\tau \frac{\Delta n^2}{\alpha^*\alpha} s_x
\left(\begin{array}{c}s_z\\0\\-s_x\end{array}\right)
 \nonumber \\ && 
+\frac{\Gamma\tau}{2}\frac{\Delta n^2}{\alpha^*\alpha} (1+s_z)
\left(\begin{array}{c}s_xs_z\\s_ys_z\\s_z^2-1\end{array}
\right)  \nonumber \\ &&
            +\frac{\Gamma\tau}{2} 
\left(\begin{array}{c}s_xs_z\\s_ys_z\\s_z^2-1\end{array}
\right) 
. 
\end{eqnarray}
The compensation term has been split into one part which compensates the
drift of the total diffusion step and the remaining term which 
modifies the effective drift. 
The nature of the diffusion term has already been discussed in section
\ref{sec:tsa}. The diffusion steps are directed towards the $s_x$ poles
of the Bloch sphere with a diffusion constant of $\Gamma(1-s_x^2)$. 
The average drift is given by
\begin{equation}
\left(\begin{array}{c}\langle \delta s_{x}\rangle \\
\langle \delta s_{y}\rangle \\
\langle \delta s_{z}\rangle \end{array}
\right) =
  + \frac{\Gamma\tau}{2}
\left(\begin{array}{c} s_xs_z^2\\s_ys_z(2+s_z)\\
-s_y^2(2+s_z)-s_x^2s_z \end{array}\right). 
\end{equation}
Figure \ref{fbdrift0} shows the drift vectors on the Bloch sphere.
The change in $s_x$ always has the same sign as $s_x$, indicating that this
drift will cause an increase in the dipole expectation value $s_x$ until an 
eigenvector with $s_x = \pm 1$ is reached. This implies that the feedback 
setup actually creates the stabilized coherent state regardless of the 
initial state. 

If the initial state is the ground state, it is sufficient to examine the 
drift term for $s_y = 0$, since neither the diffusion nor the drift will
create an $s_y \neq 0$. The drift term is then given by
\begin{equation}
\left(\begin{array}{c}\langle \delta s_{x}\rangle \\
\langle \delta s_{y}\rangle \\
\langle \delta s_{z}\rangle \end{array}
\right)_{s_y=0} =
  + \frac{\Gamma\tau}{2}s_xs_z
\left(\begin{array}{c} s_z\\0\\-s_y\end{array}\right). 
\end{equation}
This drift equation has four stationary solutions, two unstable ones at
$s_z = \pm 1$, the ground state and the excited state, and two stable
ones at $s_x=\pm 1$, the dipole eigenstates. Consequently the system state 
will be drawn towards the closest one of the two dipole eigenstates as soon
as the diffusion has moved the state away from the destabilized ground 
state.
If the phase of the local oscillator is changed, whether intentionally
or accidentally, this does create an $s_y \neq 0$ component. Since there
will always be limits to the phase stability of the local oscillator, it
is essential that the effects of such deviations are understood as well.
For $s_z =0$, the system reacts with a diffusion in the $s_x,s_y$ plane
and a drift reducing $s_z$. For small values of $s_y$, the system will
relax exponentially to $s_z = -2s_y^2$. This induces a drift in $s_y$
corresponding to
\begin{equation}
\frac{d}{dt}s_y = -2\Gamma s_y^3.
\end{equation}   
This stabilization of $s_y=0$ is very weak compared to the exponential
stability in $s_z=0$. The dynamics of the
relaxation for small deviations is approximately given by
\begin{equation}
\label{eq:approxsy}
s_y(t) = \frac{1}{\sqrt{4\Gamma t + s_y(0)^{-2}}}.
\end{equation}
Figure \ref{relaxation} shows a comparison of this relaxation with the 
exponential relaxation of $s_z$. This comparison clearly demonstrates
that the relaxation of $s_y$ may indeed be separated from the fast relaxation 
of $s_z$.  
Note that since the relaxation of $s_y$ is very slow, its effect can not be 
separated from the diffusion. The diffusion constant of $\Gamma s_y^2$ is 
actually
quite similar to the drift term of $s_y$. Consequently, the trajectory of the 
system
dynamics will be more complicated and unpredictable than suggested by 
equation (\ref{eq:approxsy}) and figure \ref{relaxation}. 

On time scales longer than $1/\Gamma$, the expectation value of the dipole
variable $s_x$ can be observed in the averages of $\Delta n$. This may serve 
as a control of the stabilization process and possibly as an additional
tool to produce the correct phase relation between the local oscillator and
the atomic system. Note that it is fairly safe to interpret small deviations
from $s_x=\pm 1$ as phase mismatches between the local oscillator and the
atomic system since the phase stability given by $s_y$ is so much weaker 
than the excitation stability given by $s_z$.  

\subsection{Implications for coupled and multi-level atomic systems}

The major part of this paper is concerned with the possibility of observing
and manipulating the dynamics of a simple atomic two state system interacting 
with the light field continuum. The theory of photon emission used may also
be extended to coupled and multi-level systems, as shown in \cite{Hof95} for 
a simple quantum beat scenario. 
One important aspect of the effects of homodyne detection of light field 
emissions from larger electronic systems, such as networks of coupled quantum
dots, is that the contribution of the system dynamics must be considered in 
more detail. In fact, the
presence of different frequencies in the system dynamics may effectively 
remove the difference between homodyne and heterodyne detection.
In quantum beat scenarios such as the ones considered in \cite{Hof95,Hof96}, 
the homodyne detection could be performed on
only one of the emission channels, with consequences for the probability 
of detecting an emission in the other channel. If the coupling strength
of the transitions is very different \cite{Hof96}, the quantum fluctuations 
coupling to
the fast transitions will induce phase fluctuations in the dynamics of the
system. Such phase fluctuations in the quantum beats of a 
system could be measured by homodyne detection. The resulting correlation
between the measurement protocols of the homodyne detection and the 
probability of an emission from a local part of the system can be predicted 
using quantum trajectory formalisms such as the one presented in this paper. 
The effective projective measurement base of an 
eight port homodyne detection which was discussed in \cite{Lui96} may 
also be applied
in place of the simple balanced homodyne detection scheme used here.
This would restore the symmetry between $s_x$ and $s_y$, increasing
the number of possible stabilization scenarios.    


\section{Interpretation and conclusions}

\label{sec:concl}

\subsection{Interpretation of the light-atom interaction}

In the usual photon detection measurement the quantum fluctuations of the
vacuum field state seem to have no effect. This impression is a result of
the particle picture interpretation associated with the photon measurement. 
If homodyne detection 
is used instead, the information obtained is mainly information about
the quantum fluctuations acting on the system, with only small contributions
from the in-phase dipole component of the atomic system. While the 
information about
the time of photon emission is lost completely, the evolution of the atomic
dipole oscillations may be reconstructed from the measurement protocols.

However, the Bloch vector does not react to the quantum fluctuations in the 
same way as it would respond to a classical driving field. The fact that no
energy may be absorbed from the quantum fluctuations of the light field vacuum
requires that the ground state is not influenced by the fluctuations.
At the same time, the corresponding excited state is much more sensitive to
the fluctuations. This asymmetry required by energy conservation may also be
understood in terms of a correlation of the atomic system and the field.
The atomic ground state is really a dressed state in which the field and
dipole fluctuations are correlated so as to preserve the state of lowest 
energy. Consequently, there is a similar correlation in the excited state,
which enhances the interaction with the quantum fluctuations. Of course,
this correlation extends only to the quantum noise. Therefore, the coherent
fields used to control the atom interact in a different manner and can never
compensate the dissipative effects. This can only be achieved by manipulating
the electromagnetic vacuum itself, for example by putting the atom into a
microcavity.

The peculiar nature of quantum states and quantum mechanical uncertainty 
is also apparent in the asymmetry of the diffusion step. If the quantum
state is considered to be an objectively real description of the atomic 
system it is difficult to 
explain the dependence of the diffusion step on the phase of the local 
oscillator used in the homodyne detection, especially since the measurement
could be performed a long time after the emission and at an arbitrary 
distance. Thus this simple balanced homodyne detection scenario
also highlights the epistemological nature of the wavefunction and the
resulting non-locality. 

The analysis of the contributions to the spontaneous emission process observed
on time scales of $1/\Gamma$ reveals a surprisingly complex structure of the
process, with contributions from quasi classical dipole radiation and 
non linear effects of the fluctuations. The interpretation of these terms is 
far from complete and reveals new challenges to our physical understanding 
presented by this fundamental quantum mechanical process.

\subsection{Conclusions}
We have presented a completely quantum mechanical theory of homodyne detection,
simulating
the time resolved observation of electromagnetic emissions from an 
atomic system at a quantum efficiency of 100\%. The theory includes 
a model of the unitary evolution of the system-field correlation and 
applies projective measurements to the results of this temporal
evolution. 
The information obtained about the dissipative dynamics of
the atomic system has been applied in feedback scenarios which
demonstrate that the dissipative dynamics can be modified by feedback to
create and stabilize excited or coherent states of the system. The dissipative
nature of the measurement can not be compensated however, and excited states
show an instability with regard to small errors in the initial preparation
on time scales of $1/\Gamma$, the natural lifetime of the excited state.
In conclusion, we have shown that homodyne detection can be a useful tool in
attempts to observe and control individual quantum systems.

\section*{Acknowledgements}
We would like to thank Ariel Liebmann and Claus Granzow for many helpful 
discussions and Howard Carmichael for valuable comments.
 

%
%
\begin{figure}
\caption{Visualization of the diffusion step on the Bloch sphere. The diffusion
is represented by lines oriented parallel to the direction of the diffusion
with a length proportional to the standard deviation of the diffusion step.
a) shows the projection into the $s_y, s_z$ plane and b) the projection into 
the $s_x,s_z$ plane.}
\label{diffusion} 
\end{figure}
\begin{figure}
\caption{Total drift on the Bloch sphere projected into the $s_y, s_z$ 
plane. Since the drift is rotationally symmetric around the $s_z$ axis, no
extra figure is given for the $s_x,s_z$ plane.}
\label{drift} 
\end{figure}
\begin{figure}
\caption{Relative contributions to the exponential decay of the atomic
excitation for $\mid \langle{\bf s}\rangle \mid = 1$.}
\label{contributions} 
\end{figure}
\begin{figure}
\caption{Visualization of the effective diffusion step including a feedback
stabilizing $\bar{s}_z = +1$. The representation is as in figure 1.}
\label{fbdiff1} 
\end{figure}
\begin{figure}
\caption{Visualization of the effective diffusion step including a feedback
stabilizing $\bar{s}_z = 0$. The representation is as in figure 1.}
\label{fbdiff0} 
\end{figure}
\begin{figure}
\caption{Effective drift on the Bloch sphere including a  feedback
stabilizing $\bar{s}_z = 0$. a) shows the projection into the $s_y, s_z$ 
plane and b) the projection into the $s_x,s_z$ plane. }
\label{fbdrift0} 
\end{figure}
\begin{figure}
\caption{Comparison of the approximate relaxation dynamics 
of $s_z(0)=0.1$ and $s_y(0)=0.1$ in the presence of a feedback 
stabilizing $\bar{s}_z = 0$.
$s_y$ does not change much while 
the exponential relaxation effectively reduces $s_z$ to zero.}
\label{relaxation} 
\end{figure}
%

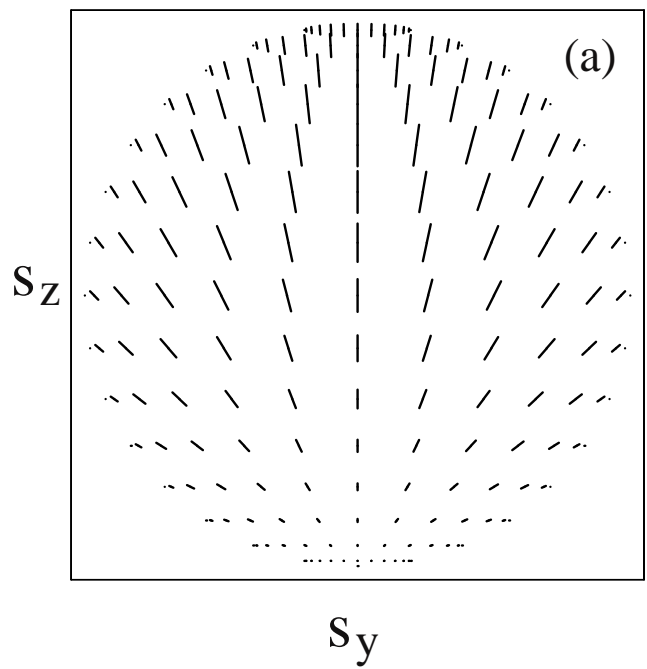
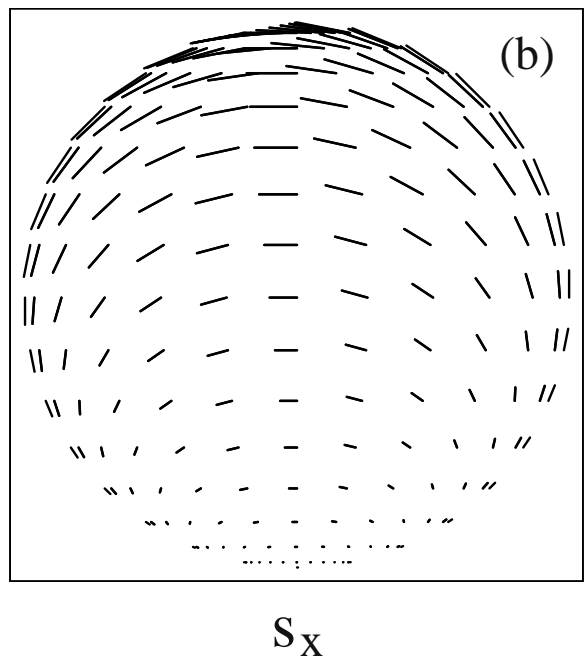
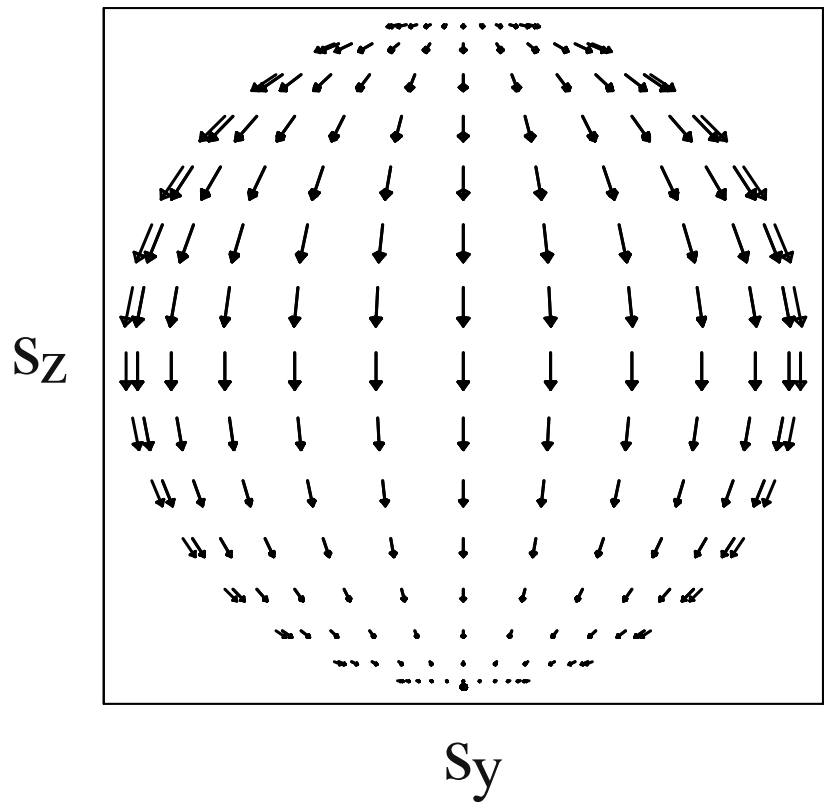
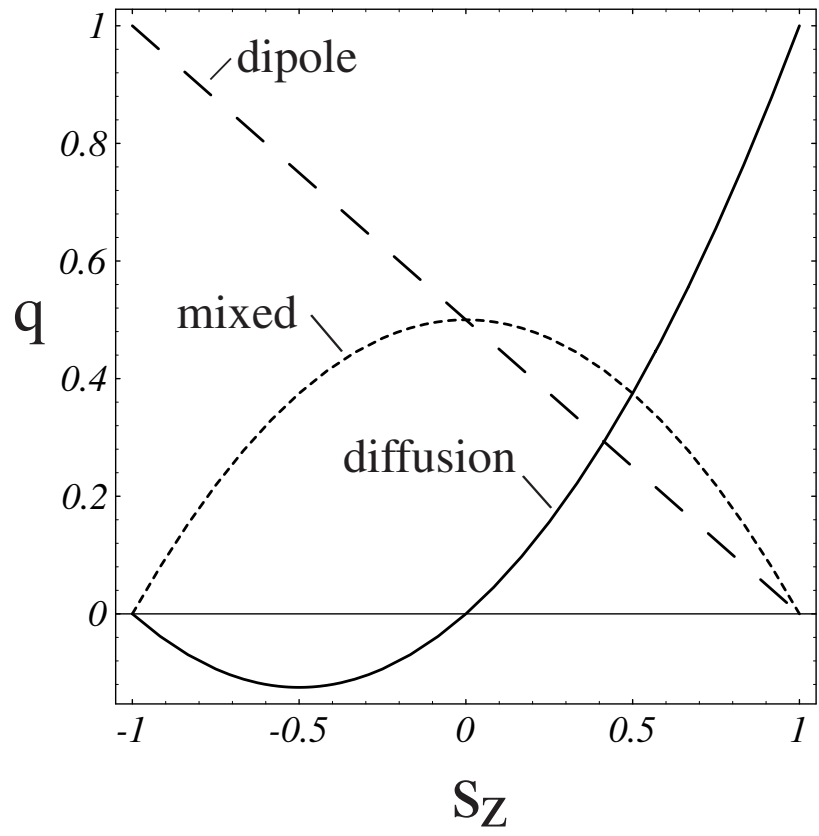
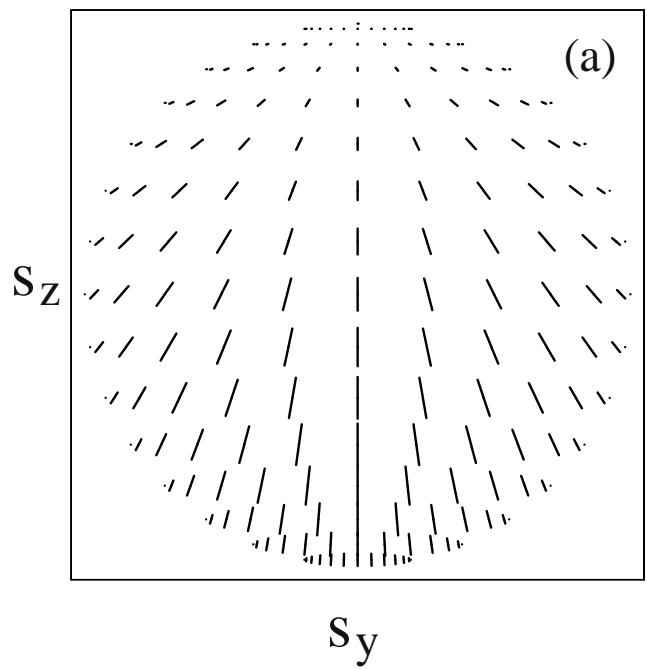
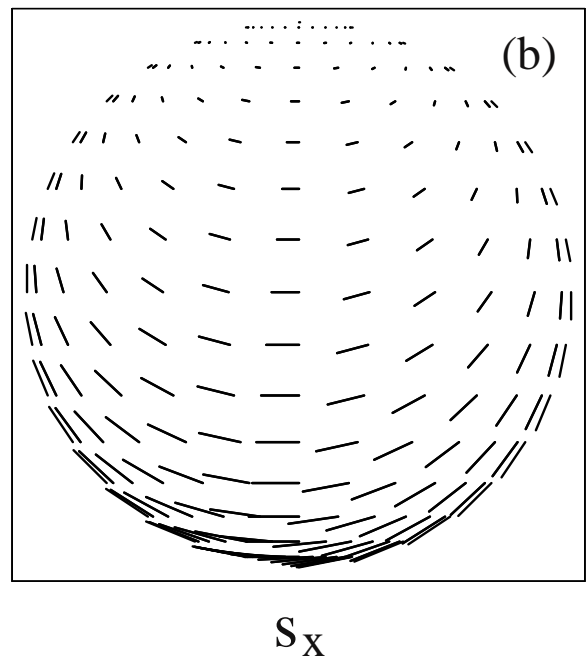
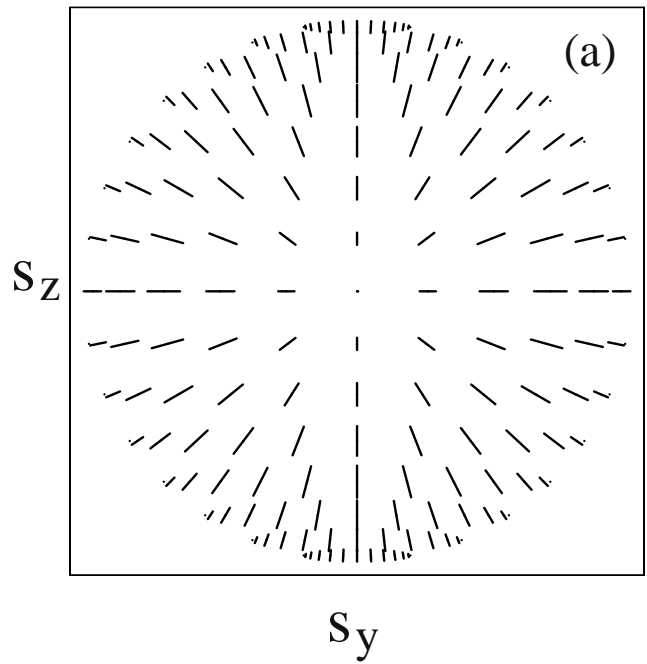
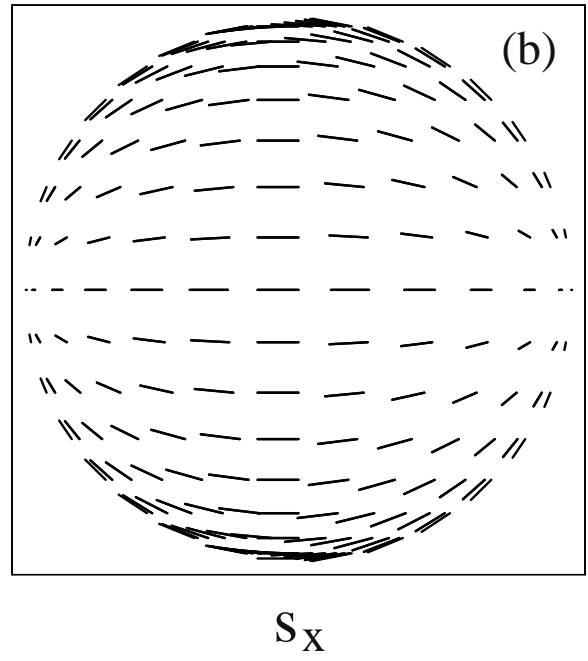
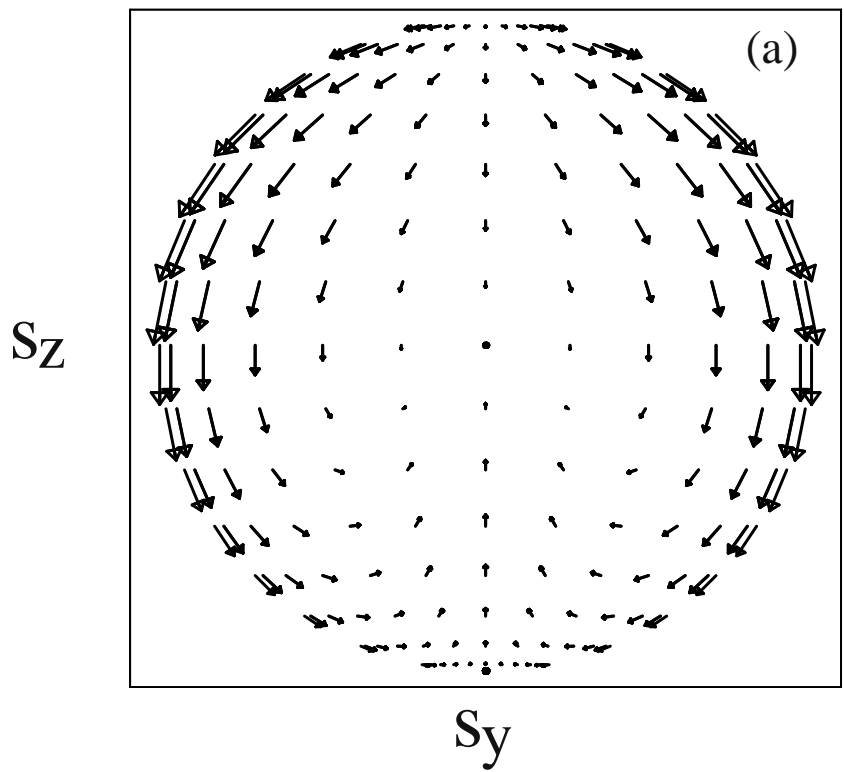
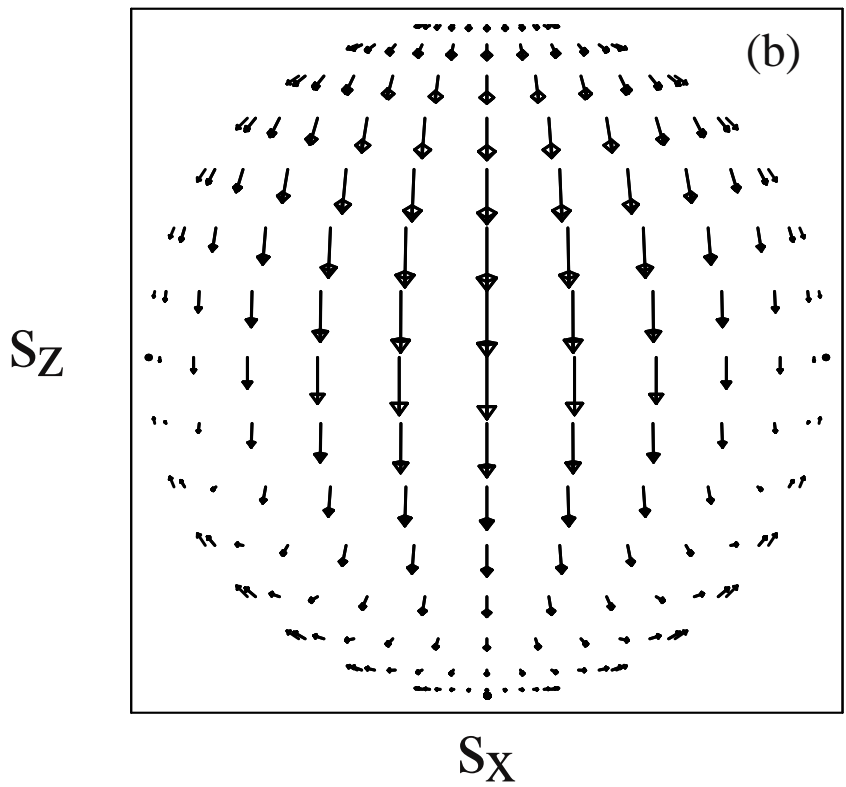
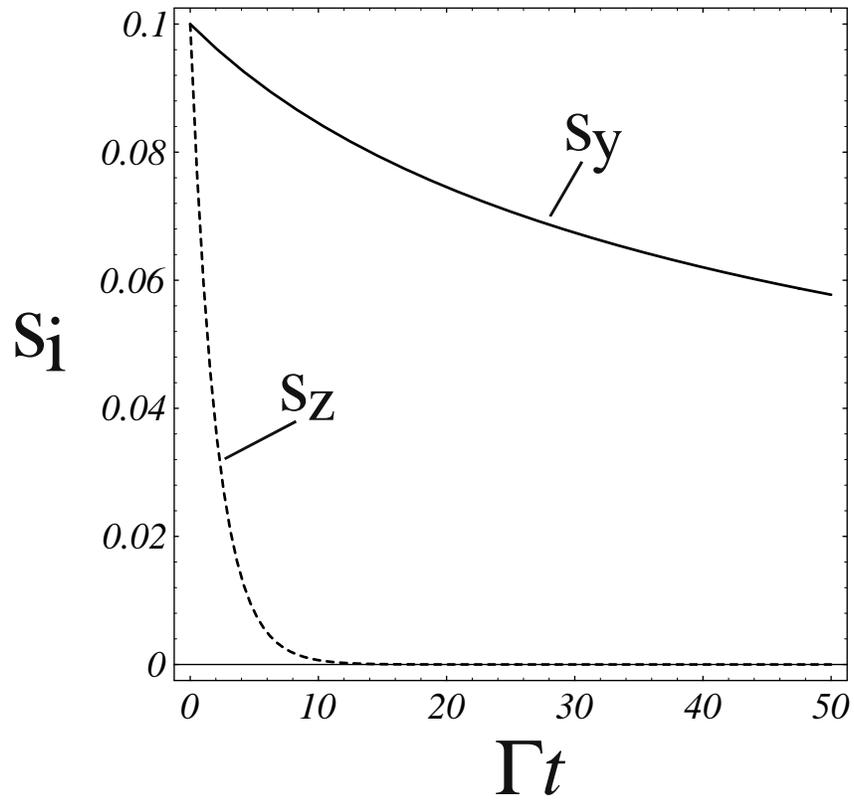
\end{document}